\documentclass[aps,tightenlines,nofootinbib,preprint,
superscriptaddress,groupedaddress,epsfig]{revtex4}

\usepackage{graphicx}

\begin{document}

\title{Annihilation Rate of Heavy $0^{++}$ P-wave Quarkonium \\
in Relativistic Salpeter Method }
\vspace{2cm}

\author{Guo-Li Wang}
\email{gl_wang@hit.edu.cn} \affiliation{Department of Physics,
Harbin Institute of Technology, Harbin, 150001, China}

\baselineskip=20pt

\vspace{2cm}

\begin{abstract}

Two-photon and two-gluon annihilation rates of P-wave scalar
charmonium and bottomonium up to third radial excited states are
estimated in the relativistic Salpeter method. We solved the full
Salpeter equation with a well defined relativistic wave function
and calculated the transition amplitude using the Mandelstam
formalism. Our model dependent estimates for the decay widths:
$\Gamma(\chi_{c0} \rightarrow 2\gamma)=3.78$ keV,
$\Gamma(\chi'_{c0} \rightarrow 2\gamma)=3.51$ keV,
$\Gamma(\chi_{b0} \rightarrow 2\gamma)=48.8$ eV and
$\Gamma(\chi'_{b0} \rightarrow 2\gamma)=50.3$ eV. We also give
estimates of total widths by the two-gluon decay rates:
$\Gamma_{tot}(\chi_{c0})=10.3$ MeV,
$\Gamma_{tot}(\chi'_{c0})=9.61$ MeV,
$\Gamma_{tot}(\chi_{b0})=0.887$ MeV and
$\Gamma_{tot}(\chi'_{b0})=0.914$ MeV.

\end{abstract}


\maketitle

\section{Introduction}

It is well known that two-photon or two-gluon annihilation rate of
heavy quarkonium $c\bar c$ or $b \bar b$ is related to the wave
function, so this process will be helpful to understand the
formalism of inter-quark interactions, and can be a sensitive test
of the potential model \cite{Godfrey}. With a replacement of the
photons by gluons, the finial state becomes two gluon state, which
will be helpful to give the information on total width of the
corresponding quarkonium.

In previous letter \cite{twophoton}, two-photon and two-gluon
annihilation rates of $0^{-+}$ pseudoscalar $c\bar c$ and $b \bar
b$ states are computed in the relativistic Salpeter method, good
agreement of our predictions with other theoretical calculations
and available experimental data is found. We also found the
relativistic corrections are important and can not be ignored. In
this letter, we extend our previous analysis to include the P-wave
$0^{++}$ scalar $c\bar c$ and $b \bar b$ states, present a
relativistic calculation of these states decaying into two photons
or two gluons.

 For the theoretical
estimates of the two-photon or two-gluon annihilation rate, we
have various methods readily available in
hand\cite{Schuler,Ma,Huang,Lakhina,Crater,Mangano,Bodwin,Patrignani,Ebert,Munz,Gupta}.
First was the non-relativistic calculation, the corresponding
decay width is related to the derivative of the non-relativistic
P-wave function at the origin, this method will cause large
uncertainty because in a full relativistic calculation the decay
width is related to the full behavior of P-wave function which can
be seen in this letter or in Ref. \cite{twophoton}. So the
relativistic corrections is very important. In recent years, many
authors try to focus on the relativistic corrections and there are
already some versions of relativistic calculation, and they give
improved results over the non-relativistic methods. In this
letter, we give yet another relativistic calculation by the
instantaneous Bethe-Salpeter method \cite{BS}, which is a full
relativistic method \cite{salp} with a well defined relativistic
form of wave function.

In previous letter \cite{twophoton}, we have pointed out that
there are two sources of relativistic corrections; one is the
correction in relativistic kinematics which appears in the decay
amplitudes through a well defined form of relativistic wave
function (related to the full behavior of wave function, not
merely related to the derivative of the wave function at origin);
the other relativistic correction comes via the relativistic
inter-quark dynamics, which requires not only a well defined
relativistic wave function but also a good relativistic formalism
to describe the interactions among quarks. The Bethe-Salpeter
equation and its instantaneous version, Salpeter equation, are
well-known tools to describe relativistic bound states. In this
letter we will solve the full Salpeter equation for $0^{++}$
state, and use the full Salpeter wave function to estimate the
annihilation decay width of quarkonium.

The form of the wave function is also important in the
calculation, since the corrections of the relativistic kinetics
come mainly through it. We begin from the quantum field theory,
analyze the parity and charge conjugation of bound state, and give
a formula for the wave function that is in a relativistic form
with definite parity and charge conjugation symmetry. Another
important thing is how to use the relativistic wave function of
bound state to obtain a relativistic transition amplitude, since a
non-relativistic transition amplitude even with a relativistic
wave function will lose the benefit of relativistic effects caused
by a relativistic wave function. The Mandelstam formalism is well
suited for the computation of relativistic transition amplitude,
and we begin with this formulism to give a formula of the
transition amplitude.

In Sec. II, we give the transition amplitude in Mandelstam
formalism and corresponding wave function with a well defined
relativistic form. In Sec. III, the full Salpeter equation is
solved, and the mass spectra and numerical value of wave function
are obtained. Then the two-photon decay width and full width of
heavy $0^{++}$ quarkonium are estimated. In Sec. III, short
discussions and a summary are also given.

\section{Theoretical Details}

According to the Mandelstam \cite{mandelstam} formalism, the
relativistic transition amplitude of a quarkonium decaying into
two photons (see figure 1) can be written as:
\begin{eqnarray}
T_{2\gamma} & = & i\sqrt{3}\; (iee_q)^2 \!
 \int \!\! \frac{d^4q}{(2\pi)^4}
              \; \mbox{tr}\; \Bigg\{ \,
    \chi(q) \bigg[\varepsilon\!\!\! /_2\, S(p_1-k_1)\,
    \varepsilon\!\!\! /_1 +
     \varepsilon\!\!\! /_1\, S(p_1-k_2)\,
    \varepsilon\!\!\! /_2 \bigg]
        \Bigg\},
\label{eq1}
\end{eqnarray}
where $k_1$, $k_2$; $\varepsilon_1$, $\varepsilon_2$ are the
momenta and polarization vectors of photons; $e_q=\frac{2}{3}$ for
charm quark and $e_q=\frac{1}{3}$ for bottom quark; $p_1$ and
$p_2$ are the momentum of constitute quark and antiquark;
$\chi(q)$ is the quarkonium Bethe-Salpeter wave function with the
total momentum $P$ and relative momentum $q$, related by
$$p_{_1}={\alpha}_{1}P+q, \;\; {\alpha}_{1}\equiv\frac{m_{1}}{m_{1}+m_{2}}~,$$
$$p_{2}={\alpha}_{2}P-q, \;\; {\alpha}_{2}\equiv\frac{m_{2}}{m_{1}+m_{2}}~,$$
where $m_1=m_2$ is the constitute quark mass of $c$ or $b$.

Since $p_{10}+p_{20}=M$, the approximation
$p_{10}=p_{20}=\frac{M}{2}$ is a good choice for the equal mass
system \cite{barbieri,keung,Huang}. Having this approximation, we
can perform the integration over $q_0$ to reduce the expression,
with the notation of Salpeter wave function
$\Psi(q)=\int\frac{dq_0}{2\pi}\chi(q)$, to
\begin{eqnarray}
T_{2\gamma}  =  \sqrt{3}\; (ee_q)^2 \!
 \int \!\! \frac{d\vec{q}}{(2\pi)^3}
              \; \mbox{tr}\; \Bigg\{ \,
    \Psi(\vec q) \bigg[\varepsilon\!\!\! /_2\, \frac{1}{\not\! p_1-\not\! k_1-m_1}\,
    \varepsilon\!\!\! /_1 +
     \varepsilon\!\!\! /_1\, \frac{1}{\not\! p_1-\not\! k_2-m_1}\,
    \varepsilon\!\!\! /_2 \bigg]
        \Bigg\}.
\label{eq2}
\end{eqnarray}
With this relativistic amplitude, the two photon decay width can
be written as
   \begin{eqnarray}
   \Gamma_{2\gamma}=\frac{{T^{~2}_{2\gamma}}}{16\,\pi M}.
   \label{eq6}
   \end{eqnarray}

The general form for the relativistic wave function of scalar
state $J^{PC}=0^{++}$ can be written as $8$ terms constructed by
momentum $P$, $q$ and Dirac matrix $\gamma$, because of the
approximation of instantaneous, $4$ terms with $P\cdot q$ become
zero, with further constraint from Salpeter equation, the
relativistic Salpeter wave function $\Psi(\vec q)$ for $0^{++}$
state with a definite parity $(+)$ and charge conjugation $(+)$
can be written as:
\begin{equation}
\Psi(\vec{q})= \varphi_1(\vec{q}) \, {\not\! \vec q}
+\varphi_2(\vec{q})\,\gamma_{0}\,{\not\! \vec q}-\frac{ { \vec
q}^2}{m_1}\, \varphi_1(\vec{q})\;. \label{eq3}
\end{equation}
The wave function $\varphi_1(\vec{q})$, $\varphi_2(\vec{q})$ and
bound state mass $M$ can be obtained by solving the full Salpeter
equation with the constituent quark mass as input, and they should
satisfy the normalization condition:
\begin{equation}\int\frac{d{\vec{q}}}{(2\pi)^3}
\frac{8\,\omega_{1}\,{\vec{q}}^2}{m_{1}}\,\varphi_1
({\vec{q}})\,\varphi_2({\vec{q}}) =2M~,\label{eq4}
\end{equation}
where $\omega_{1}=\sqrt{m_{1}^{2}+{\vec q}^{2}}$.

The two gluon decay width of quarkonium can be easily obtained
from the two photon decay width, with a simple replacement in
the photon decay width formula
\begin{equation}
{e_q}^4{\alpha}^2\longrightarrow \frac{2}{9}\,{\alpha_s}^2.
\label{eq7}
\end{equation}

\begin{figure}
\begin{picture}(250,130)(200,400)
\put(0,0){\includegraphics{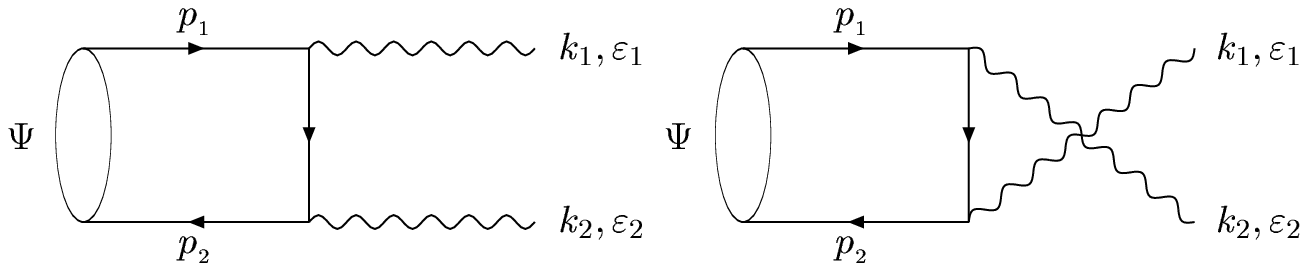}}
\end{picture}
\caption{Two-photon annihilation diagrams of the quarkonium. }
\end{figure}

\section{Numerical Results and Discussions}

We will not show the details of Solving the full Salpeter
equation, only give the final results, interested readers can find
the the detail technique in Ref.~\cite{cskimwang}.

When solving the full Salpeter equation, we choose a
phenomenological Cornell potential, there are some parameters in
this potential including the constitute quark mass and one loop
running coupling constant. Since we lack data of $0^{++}$ $c\bar
c$ states to determine these parameters, we choose the same values
as in Ref.~\cite{cskimwang} which were obtained by fitting the
mass spectra for $0^{-}$ states, but only vary parameter $V_0$ to
fit the mass of $\chi_{c0}$, so the following parameters are
adopted: $a=e=2.7183$, $\alpha=0.06$ GeV, $V_0=-0.566$ GeV,
$\lambda=0.2$ GeV$^2$, $\Lambda_{QCD}=0.26$ GeV and
 $m_c=1.7553$ GeV.
With this parameter set, we solve the full Salpeter equation and
obtain the mass spectra shown in Table I. With the obtained wave
function and Eq. (\ref{eq6}), we calculate the two-photon decay
width of $c\bar c$ $0^{++}$ states, the results are shown in Table
I. To give the numerical analysis of two gluons decay, we need to
fix the value of the renormalization scale $\mu$ in
$\alpha_s(\mu)$. In the case of charmonium, we choose the charm
quark mass $m_c$ as the energy scale and obtain the coupling
constant $\alpha_s(m_c)=0.36$ \cite{cskimwang}. The corresponding
total decay width are also listed in Table I.

For the case of $b \bar b$ system, to determine the parameters, we
fit the available masses $M_{\chi_{b0}}=9.8599$ GeV,
$M_{\chi'_{b0}}=10.2321$ GeV,  and we also fit the mass of
$\eta_b$ at $9.364$ GeV \cite{twophoton}. So the parameters are
$V_0=-0.553$ GeV, $\Lambda_{\rm QCD}=0.20$ and $m_b=5.13$ GeV,
other parameters are same as in the case of $c \bar c$. With this
set of parameters, the coupling constant at scale of bottom quark
mass is $\alpha_s(m_b)=0.232$. The corresponding mass spectra,
two-photon and total decay widths are shown in Table II.

\begin{table*}[hbt]
\setlength{\tabcolsep}{0.5cm} \caption{\small Two-photon decay
width and total width of P-wave $0^{++}$ charmonium states, where
the total width is estimated by the two-gluon decay width
$\Gamma_{tot}\simeq\Gamma_{2g}$.} \label{tab1}
\begin{tabular*}{\textwidth}{@{}c@{\extracolsep{\fill}}cccc}
 \hline \hline
{\phantom{\Large{l}}}\raisebox{+.2cm}{\phantom{\Large{j}}}
&Mass~~GeV&$\Gamma_{2\gamma}$~~keV &$\Gamma_{tot}$~~MeV\\
\hline\hline
{\phantom{\Large{l}}}\raisebox{+.2cm}{\phantom{\Large{j}}}
$\chi_{c0}(1P)$~&~3.4159~&~3.78 &10.3\\
{\phantom{\Large{l}}}\raisebox{+.2cm}{\phantom{\Large{j}}}
$\chi^{'}_{c0}(2P)$~&~3.8311~&~3.51 &9.61\\
{\phantom{\Large{l}}}\raisebox{+.2cm}{\phantom{\Large{j}}}
$\chi^{''}_{c0}(3P)$~&~4.1324~&~2.86 &7.83\\
{\phantom{\Large{l}}}\raisebox{+.2cm}{\phantom{\Large{j}}}
$\chi^{'''}_{c0}(4P)$~&~4.3694~&~2.42 &6.62
 \\\hline\hline
\end{tabular*}
\end{table*}

\begin{table*}[hbt]
\setlength{\tabcolsep}{0.5cm} \caption{\small Two-photon decay
width and total width of P-wave $0^{++}$ bottomonium states, where
the total width is estimated by the two-gluon decay width
$\Gamma_{tot}\simeq\Gamma_{2g}$.} \label{tab2}
\begin{tabular*}{\textwidth}{@{}c@{\extracolsep{\fill}}cccc}
 \hline \hline
{\phantom{\Large{l}}}\raisebox{+.2cm}{\phantom{\Large{j}}}
&Mass~~GeV&$\Gamma_{2\gamma}$~~eV &$\Gamma_{tot}$~~MeV\\
\hline\hline
{\phantom{\Large{l}}}\raisebox{+.2cm}{\phantom{\Large{j}}}
$\chi_{b0}(1P)$~&~9.8601~&~48.8 &0.887\\
{\phantom{\Large{l}}}\raisebox{+.2cm}{\phantom{\Large{j}}}
$\chi^{'}_{b0}(2P)$~&~10.2239~&~50.3 &0.914\\
{\phantom{\Large{l}}}\raisebox{+.2cm}{\phantom{\Large{j}}}
$\chi^{''}_{b0}(3P)$~&~10.4970~&~44.7 &0.813\\
{\phantom{\Large{l}}}\raisebox{+.2cm}{\phantom{\Large{j}}}
$\chi^{'''}_{b0}(4P)$~&~10.7192~&~41.9 &0.761
 \\\hline\hline
\end{tabular*}
\end{table*}

\begingroup
\squeezetable
\begin{table*}[hbt]
\setlength{\tabcolsep}{0.5cm} \caption{\small Recent theoretical
and experimental results of two-photon decay width and total
width.} \label{tab3}
\begin{tabular*}{\textwidth}{@{}c@{\extracolsep{\fill}}ccccccc}
 \hline \hline
{\phantom{\Large{l}}}\raisebox{+.2cm}{\phantom{\Large{j}}}
&$\Gamma^{\chi_c}_{2\gamma}$~keV ~&$\Gamma^{\chi_c}_{tot}$~MeV~&
$\Gamma^{\chi^{'}_c}_{2\gamma}$~keV~&
$\Gamma^{\chi_b}_{2\gamma}$~eV~ &$\Gamma^{\chi_b}_{tot}$~MeV~&
$\Gamma^{\chi^{'}_b}_{2\gamma}$~eV~\\\hline\hline
{\phantom{\Large{l}}}\raisebox{+.2cm}{\phantom{\Large{j}}}
Ours~&~3.78 &10.3&3.51&48.8&0.887&50.3\\
{\phantom{\Large{l}}}\raisebox{+.2cm}{\phantom{\Large{j}}}
Crater\cite{Crater} ~&~3.96,~3.34 &&&&&\\
 {\phantom{\Large{l}}}\raisebox{+.2cm}{\phantom{\Large{j}}}
Gupta \cite{Gupta}  ~&~6.38&13.44&&80&2.15&\\
{\phantom{\Large{l}}}\raisebox{+.2cm}{\phantom{\Large{j}}}
Huang \cite{Huang}~&~3.72$\pm$1.11 &12.5$\pm$3.2&&&\\
{\phantom{\Large{l}}}\raisebox{+.2cm}{\phantom{\Large{j}}}
Ebert \cite{Ebert} ~&~2.9 &&1.9&38&&29&\\
{\phantom{\Large{l}}}\raisebox{+.2cm}{\phantom{\Large{j}}}
 M\"unz \cite{Munz} ~&~1.39$\pm$0.16 &&1.11$\pm$0.13&24$\pm$3&&26$\pm$2\\
 {\phantom{\Large{l}}}\raisebox{+.2cm}{\phantom{\Large{j}}}
 E835 \cite{E835} ~&~ &9.8$\pm$1.0$\pm$0.1&&&&\\
 {\phantom{\Large{l}}}\raisebox{+.2cm}{\phantom{\Large{j}}}
 CLEO \cite{CLEO} ~&~3.76$\pm$0.65$\pm$0.41$\pm$1.69 &&&&&\\
 {\phantom{\Large{l}}}\raisebox{+.2cm}{\phantom{\Large{j}}}
  PDG \cite{PDG} ~&~2.87$\pm 0.54$ &10.4$\pm 0.7$&&&&\\
  {\phantom{\Large{l}}}\raisebox{+.2cm}{\phantom{\Large{j}}}
  BES \cite{BES} ~&~ &12.6$^{+1.5+0.9}_{-1.6-1.1}$&&&&\\
 \hline\hline
\end{tabular*}
\end{table*}
\endgroup

We compare our predictions with recent other theoretical
relativistic calculations and experimental results in Table III.
Our results of $\Gamma^{\chi_{c0}}_{2\gamma}$ agree with the
estimates of Refs. \cite{Huang,Crater,CLEO}. All the values of
$\Gamma^{\chi_{c0}}_{tot}$ listed in the table consist with each
other. But there are large discrepancy in the results of
$\chi^{'}_{c0}$, $\chi_{b0}$ and $\chi^{'}_{b0}$ from different
methods, more theoretical calculations and experimental
measurements are required for these charnels.

To show the important of the relativistic corrections, we give a
estimate of non-relativistic results using the same method and
with same parameters. The non-relativistic wave function can be
written as $\varphi_1(\vec{q}) (1 +\gamma_{0}){\not\! \vec q}$,
with this wave function we solve the positive part of Salpeter
equation, and delete the contribution of negative part. The
non-relativistic normalization condition is $\int d{\vec{q}}\,
8\,\varphi_1 ({\vec{q}})^2 =2M\,{(2\pi)^3}$. When calculate the
transition amplitude, we only keep the lowest order contribution
and delete all the higher order contributions by counting the
exponential number of $|\vec q|$, the corresponding results are
show in Table IV. One can see that the relativistic corrections
are important especially for the $c \bar c$ system.

\begin{table*}[hbt]
\setlength{\tabcolsep}{0.5cm} \caption{\small Mass spectra and
Two-photon decay width with and without relativistic corrections,
where the "non-rel" means the non-relativistic results. The decay
width is in unit KeV for $c\bar c$ system and eV for $b\bar b$
system.} \label{tab4}
\begin{tabular*}{\textwidth}{@{}c@{\extracolsep{\fill}}ccccc}
 \hline \hline
{\phantom{\Large{l}}}\raisebox{+.2cm}{\phantom{\Large{j}}}
&Mass~~GeV&non-rel&$\Gamma_{2\gamma}$ &non-rel\\
\hline\hline
{\phantom{\Large{l}}}\raisebox{+.2cm}{\phantom{\Large{j}}}
$\chi_{c0}(1P)$~&~3.4159&~3.3931&~3.78 &5.85\\
{\phantom{\Large{l}}}\raisebox{+.2cm}{\phantom{\Large{j}}}
$\chi^{'}_{c0}(2P)$~&~3.8311&~3.8267&~3.51 &5.47\\
{\phantom{\Large{l}}}\raisebox{+.2cm}{\phantom{\Large{j}}}
$\chi^{''}_{c0}(3P)$~&~4.1324&~4.1543&~2.86 &4.61\\
{\phantom{\Large{l}}}\raisebox{+.2cm}{\phantom{\Large{j}}}
$\chi^{'''}_{c0}(4P)$~&~4.3694~&4.4205&~2.42 &3.94\\
{\phantom{\Large{l}}}\raisebox{+.2cm}{\phantom{\Large{j}}}
$\chi_{b0}(1P)$~&~9.8601&~9.8471&~48.8 &58.3\\
{\phantom{\Large{l}}}\raisebox{+.2cm}{\phantom{\Large{j}}}
$\chi^{'}_{b0}(2P)$~&~10.2239&~10.2129&~50.3 &59.9\\
{\phantom{\Large{l}}}\raisebox{+.2cm}{\phantom{\Large{j}}}
$\chi^{''}_{b0}(3P)$~&~10.4970&~10.4906&~44.7 &54.0\\
{\phantom{\Large{l}}}\raisebox{+.2cm}{\phantom{\Large{j}}}
$\chi^{'''}_{b0}(4P)$~&~10.7192&~10.7196&~41.9 &48.3
 \\\hline\hline
\end{tabular*}
\end{table*}

We comment that in this work we did not include the QCD radiative
correction because we focus mainly on the relativistic
corrections, though it is no doubt that the QCD correction is very
important and interesting topic.

In summary, by solving the relativistic full Salpeter equation
with a well defined form of wave function, we estimate two-photon
decay rates: $\Gamma(\chi_{c0} \rightarrow 2\gamma)=3.78$ keV,
$\Gamma(\chi'_{c0} \rightarrow 2\gamma)=3.51$ keV,
$\Gamma(\chi_{b0} \rightarrow 2\gamma)=48.8$ eV and
$\Gamma(\chi'_{b0} \rightarrow 2\gamma)=50.3$ eV, and the total
decay widths: $\Gamma_{tot}(\chi_{c0})=10.3$ MeV,
$\Gamma_{tot}(\chi'_{c0})=9.61$ MeV,
$\Gamma_{tot}(\chi_{b0})=0.887$ MeV and
$\Gamma_{tot}(\chi'_{b0})=0.914$ MeV.\\


\acknowledgements

This work was supported in part by the National Natural Science
Foundation of China (NSFC) under Grant No. 10675038.
\\

\end{document}